\documentclass[conference]{IEEEtran}
\IEEEoverridecommandlockouts
\usepackage{cite}
\usepackage{amsmath,amssymb,amsfonts}
\usepackage{algorithmic}
\usepackage{textcomp}
\usepackage{xcolor}
\usepackage[pdftex]{graphicx}
\usepackage{epstopdf}

\usepackage{amsmath,amssymb,enumerate} 
\usepackage{bm} 
\usepackage{indentfirst} 
\usepackage{empheq} 
\usepackage{here} 

\def\BibTeX{{\rm B\kern-.05em{\sc i\kern-.025em b}\kern-.08em
    T\kern-.1667em\lower.7ex\hbox{E}\kern-.125emX}}

\begin{document}

\title{A New Model of Flaming Phenomena\\ in Online Social Networks that Considers Resonance Driven by External Stimuli\\ }

\author{\IEEEauthorblockN{Tomoya Kinoshita}
\IEEEauthorblockA{\textit{Graduate School of Systems Design} \\
\textit{Tokyo Metropolitan University}\\
Hino, Tokyo 191-0065, Japan \\
{\tt kinoshita-tomoya@ed.tmu.ac.jp}}
\and
\IEEEauthorblockN{Masaki Aida}
\IEEEauthorblockA{\textit{Graduate School of Systems Design} \\
\textit{Tokyo Metropolitan University}\\
Hino, Tokyo 191-0065, Japan\\
{\tt aida@tmu.ac.jp}}
}

\maketitle

\begin{abstract}
The explosive user dynamics represented by flaming phenomena in online social networks can sometimes negatively influence lives in the real world.
To take measures against online flaming phenomena promptly, it is necessary to model its defining characteristics.
Based on the oscillation model that describes user dynamics on networks, previous work has revealed that online flaming arises when some eigenvalues of the matrix expressing network structure are non-real numbers.
This paper considers the network resonance driven by periodic external stimuli and proposes a flaming model that posits flaming even if all the matrix's eigenvalues are real numbers. 
Also, we describe a theoretical framework for observing the omen of online flaming to trigger preventive measures.
\end{abstract}

\begin{IEEEkeywords}
online social networks, user dynamics, resonance, flaming
\end{IEEEkeywords}

\section{Introduction}
\label{sec.1}
The spread of social media services continues to progress rapidly, and society is actively exchanging a lot of information through them. 
Social Networking Services (SNS), in particular, are used by many people as tools to promote communication and have become an indispensable part of daily life. 
While the information network improves the quality of our lives, we need to pay attention to the fact that it can negatively impact us.
The explosive user dynamics on networks, represented by online flaming phenomena, can negatively influence individuals' or companies' social activities in the real world. 
Typical examples are the psychological distress of cyberbullying and loss of business by scaremongering.
Since these cases occur frequently, we need to consider countermeasures to be online flaming.
Although taking measures for each case is generally, this takes a lot of time and effort.
To take countermeasures more quickly and minimize the damage, it is necessary to clarify the impact of user activities on the occurrence of online flaming.
Therefore, it is important to understand user dynamics in online social networks (OSNs) and discover the basic mechanism of online flaming \cite{aidabook,aida2}.

One of the factors behind flaming in OSNs is the dissemination of information by mass media.
Such dissemination can trigger the sharing or discussion of information among users.  
In this paper, we call the information distributed by mass media external stimuli.
Online flaming caused by external stimuli is not taken into account in the conventional model.
Given the prevalence of mass media, it is essential to understand how it can lead to online flaming and develop a model that can provide countermeasures.

This paper uses the oscillation model \cite{aidabook,aida2} to describe user dynamics in OSNs. 
This model is based on the wave equation on networks, which describes phenomena inter-user influences propagate via OSN at a finite speed. 
The oscillation model clarifies that online flaming arises when some eigenvalues of the Laplacian matrix expressing network structure are non-real numbers.

In this paper, we propose a model of online flaming caused by the resonance driven by external stimuli; online flaming is possible even if all the Laplacian matrix's eigenvalues describing network structure are real numbers.
In the forced oscillation of a physical harmonic oscillator, applying a periodic external force whose angular frequency is close to the oscillator's eigenfrequency causes a phenomenon called resonance.
We can apply this mechanism to the oscillation modes in the oscillation model. 
Therefore, we need to elucidate the online flaming structure caused by the resonance in user dynamics in OSNs due to the injection of external stimuli. 

The rest of this paper is organized as follows.
Sec.~\ref{sec.2} introduces key related work.
Sec.~\ref{sec.3} explains the theoretical model used in this paper to describe user dynamics in OSNs.
Sec.~\ref{sec.4} proposes the model of online flaming that encompasses the resonance created by external stimuli and confirms its effectiveness through numerical experiments. 
Sec.~\ref{sec.5} provides the theoretical framework for identifying the omen of flaming and validates it experimentally.
Finally, we state conclusions and future work in Sec.~\ref{sec.6}.

\section{Related Work}
\label{sec.2}
Explosive user dynamics, which include online flaming phenomena, can cause the individual or social unrest.
Therefore, many studies have tackled to understand the user dynamics in OSNs from different viewpoints.
In \cite{centola}, the effects of network structure on the spread of behaviors in OSNs were investigated.
They reported that behaviors spread farther and faster across clustered-lattice networks than across corresponding random networks.
In \cite{cannarella}, the emergence and abandonment dynamics of user activity in OSNs could be explained using epidemiological models.
The authors proposed the irSIR model, which is a modification of the traditional SIR model, and applied it to real data to predict a particular SNS's future.
In \cite{Olfati}, consensus problems on networks of dynamic agents with fixed and switching topologies were discussed by analyzing three cases.
It was revealed that balanced digraphs play a key role in addressing average-consensus problems.
Thus, the dynamics of online social networks have been investigated from the relationship between network structure and information dissemination\cite{centola,isham}; others have used epidemiological models to describe information diffusion\cite{nekovee,zhao,cannarella}, while others have addressed the consensus problems\cite{Olfati,altafini}.
The above models cannot describe the explosive user dynamics including online flaming. 

Many theoretical and empirical types of studies on online flaming have also been reported.
In \cite{Moor}, online flaming on YouTube was studied using surveys of YouTube users.
The authors found that most users did not think of online flaming as a problem that impacted themselves.
Also, online flaming is often meant to express disagreement or to respond to perceived offense by others.
In \cite{Hwang}, the neutralization theory, the theory of planned behavior, was integrated with motivational theory to develop a theoretical model to understand better online flaming in virtual communities.
The results of online questionnaires indicated that acceptability, enjoyment, subjective norms, and low self-control are significant factors influencing online flaming.
In \cite{Hmielowski}, a communication process model was proposed that focused on what conditions lead people to engage in aggressive online communication behaviors.
Their results demonstrated that the intention to flame is greater among those that evidence high verbal aggression levels.
We note that there are various definitions of online flaming; some define it as showing hostility through offensive words and insults\cite{Moor}; others define it from the aspect of human communicative behavior\cite{Sullivan}, while others define it as messages that project attributes such as aggression, intimidation, and sarcasm\cite{Turnage}.
However, to solve online flaming more completely and quickly, understanding the behavior of users who cause online flaming is necessary.

The interactions between users generate user dynamics in OSNs.
Unfortunately, the interactions are so complex that it is so difficult to describe them in detail.
Therefore, the oscillation model was proposed to express the interactions between users in OSNs \cite{aida2,aida1}.
The oscillation model illuminates the relationship between network structure and explosive user dynamics \cite{aida2,aida3}.
Specifically, the potential for explosive user dynamics exists when some Laplacian matrix's eigenvalues are non-real numbers.
Some countermeasures to flaming in networks have been proposed, such as manipulating the structure of directed links \cite{aida3} or adjusting the damping coefficient in the oscillation model \cite{nagatani1}.
In \cite{nagatani2}, theoretical and experimental studies on the onset of online flaming were detailed.
However, these models considered only the network structure in elucidating explosive user dynamics, including online flaming, and so ignored the impact of external stimuli.
In this paper, we consider the resonance created in the network by external stimuli for a fuller understanding of online flaming.
The resonance of oscillation dynamics in OSNs was applied to investigate the structure of OSNs \cite{furutani}, but its relationship with online flaming was not considered.
We describe the impact of information spread by mass media on the user dynamics in OSNs and clarify the mechanism of online flaming from an engineering perspective.
Also, we introduce a theoretical framework for discerning the omen of online flaming with the aim of triggering effective and timely preventive measures.

\section{Theoretical Model of User Dynamics in Online Social Networks}
\label{sec.3}
In this section, based on \cite{aida2,aida1}, we describe a theoretical model of user dynamics in OSNs.
First, we introduce the Laplacian matrix \cite{spectral} to express the structure of networks.
Next, we describe the oscillation model on the networks.
Finally, we explain resonance in oscillation dynamics on networks.
It plays an important role in developing a comprehensive model of online flaming.

\subsection{Laplacian Matrix}
\label{subsec.3A}
Let $\mathcal{G}(V,\, E)$ be a directed graph representing the structure of OSNs with $n$ nodes, where $V$ is the set of nodes, and $E$ is the set of directed links. 
Also, let $w_{ij} > 0$ be the weight of the directed link $(i \rightarrow j) \in E$ from node $i$ to node $j$.
Here, nodes and links represent users and connections between them, respectively.
The (weighted) adjacency matrix $\bm{\mathcal{A}} :=[\mathcal{A}_{ij}]_{1\le i,j\le n}$ is an $n \times n$ matrix defined as
\begin{align}
  \mathcal{A}_{ij} :=
  \begin{cases}
    w_{ij},&(i \rightarrow j) \in E,\\
    0,&(i \rightarrow j) \notin E.
  \end{cases}
\end{align}
For the weighted out-degree $d_i := \sum_{j=1}^{n} \mathcal{A}_{ij}$ of node $i$, the degree matrix $\bm{\mathcal D}$ is an $n \times n$ matrix defined as
\begin{align}
  \bm{\mathcal{D}} := \text{diag}(d_1,\,d_2,\,\ldots,\,d_n).
\end{align}
The Laplacian matrix $\bm{\mathcal{L}}$ of the (weighted) directed graph is defined as
\begin{align}
  \bm{\mathcal{L}} := \bm{\mathcal{D}} - \bm{\mathcal{A}}.
  \label{2.4}
\end{align}
In general, the Laplacian matrix $\bm{\mathcal{L}}$ is asymmetric since the structure of OSNs is a directed graph.

We assume all the eigenvalues of the Laplacian matrix $\bm{\mathcal{L}}$ are real. 
This means a situation that user dynamics do not diverge unless an external force is injected. 
It is known that the minimum eigenvalue of $\bm{\mathcal{L}}$ is zero.
Therefore, $\bm{\mathcal{L}}$ has the left eigenvector ${}^t\!\bm{m} = (m_1,\,m_2,\, \ldots ,\,m_n)$ associated with eigenvalue zero; that is,
\begin{align}
{}^t\!\bm{m}\,\bm{\mathcal{L}} = (0,\,0,\,\ldots,0).
\end{align}
If component $m_i > 0$ of ${}^t\!\bm{m}$ and the weight of directed link $w_{ij} > 0$ satisfy
\begin{align}
m_i\,w_{ij} = m_j\,w_{ji},
\label{2a}
\end{align}
for all node pairs $i$ and $j$, the directed graph is symmetrizable.
We denote the Laplacian matrix of the symmetrizable directed graph by $\bm{\mathcal{L}}_0$.

We introduce the scaled Laplacian matrix $\bm{S}_0$ defined as\begin{align}
\bm{S}_0 := \bm{M}^{+\frac{1}{2}}\,\bm{\mathcal{L}}_0\,\bm{M}^{-\frac{1}{2}},
\end{align}
where $\bm{M} := \text{diag}(m_1,\,m_2, \ldots ,\, m_n)$.
Note that $\bm{S}_0$ is a real symmetric matrix, and $\bm{S}_0$ has the same eigenvalues as $\bm{\mathcal{L}}_0$.
Also, we can obtain the eigenvector of $\bm{S}_0$ by multiplying $\bm{M}^{+\frac{1}{2}}$ from the left to the eigenvector of $\bm{\mathcal{L}}_0$. 
So, we can analyze the symmetrizable directed graph more easily by using the scaled Laplacian matrix $\bm{S}_0$ instead of $\bm{\mathcal{L}}_0$.

We assume the structure of OSNs is a symmetrizable directed graph. 
This assumption is too strong to describe actual OSNs. 
However, symmetrizable directed graphs are known to be the safest conditions for the occurrence of online flaming and are a desirable premise for discussing the occurrence of online flaming associated with the injection of external stimuli.

\subsection{Oscillation Model on Symmetrizable Directed Networks}
\label{subsec.3B}
To describe the user dynamics on networks, let us consider an interaction model between users as simple and universal as possible.
Let $x_i(t)$ be the state of node $i$ at time $t$, and each node is subjected to a force from each adjacent node. 
The force is acting toward the difference of the states of the own node and the adjacent node to be $0$, and its strength is proportional to the difference.
That is, node $i$ is subjected to the restoring force that is represented as 
\begin{align}
  f_{i\rightarrow j} = -w_{ij}(x_i(t) - x_j(t)),
\end{align}
where $f_{i\rightarrow j}$ is the force acting on node $i$ from adjacent node $j$; $w_{ij}$ is a positive constant.

Based on the above interaction model, we introduce a forced oscillation model on networks.
Considering that information spreads in OSNs at a finite speed, external stimuli should be represented as a wave in the wave equation-based model.
Therefore, we consider the situation that we impose a periodic external stimulus with angular frequency $\omega$ and amplitude $F$ on a certain node, $j$.
Also, each node is subjected to a damping force that is proportional to its own velocity.
The equation of motion of the node state vector $\bm{x}(\omega,\,t) := {}^t\!(x_1(\omega,\,t),\,x_2(\omega,\,t),\,\ldots,\,x_n(\omega,\,t))$ for the forced oscillation of a symmetrizable directed graph can be written by using its Laplacian matrix $\bm{\mathcal L}_0$ as follows:
\begin{align}
  \frac{\partial^2\bm{x}(\omega,t)}{\partial t^2} +
  {\gamma}\frac{\partial\bm{x}(\omega,t)}{\partial t} + \bm{\mathcal L}_0\,\bm{x}(\omega,t)
  = F\cos(\omega t)\,\bm{1}_{\{j\}},
  \label{F_O_L0}
\end{align}
where $\gamma \geq 0$ is the damping coefficient and $\bm{1}_{\{j\}}$ is an $n$-dimensional vector whose $j$-th component is $1$ and all others are $0$.
By multiplying this equation by $\bm{M}^{+\frac{1}{2}}$ from the left, we obtain 
\begin{align}
\frac{\partial^2\bm{y}(\omega,t)}{\partial t^2} +
{\gamma}\frac{\partial\bm{y}(\omega,t)}{\partial t} + \bm{S}_0\,\bm{y}(\omega,t)
= \sqrt{m_j}\,F\cos(\omega t)\,\bm{1}_{\{j\}},
\label{F_O}
\end{align}
where $\bm{y}(\omega,\,t) = \bm{M}^{+\frac{1}{2}}\,\bm{x}(\omega,\,t)$.

Let $\lambda_{\mu}\,(\mu = 0,\,1,\,\ldots,\,n-1)$ be the eigenvalue of $\bm{S}_0$.
Also, we can choose eigenvector $\bm{v}_{\mu}$ of $\bm{S}_0$ associated with $\lambda_{\mu}$ as the orthonormal basis.
We expand $\bm{y}(\omega,\,t)$ and $\bm{1}_{\{j\}}$ by using $\bm{v}_{\mu}$ as follows:
\begin{align}
  \bm{y}(\omega,\,t) = \sum_{{\mu}=0}^{n-1} a_{\mu}({\omega},\,t)\,\bm{v}_{\mu},\quad
  \bm{1}_{\{j\}} = \sum_{{\mu}=0}^{n-1} b_{\mu}\,\bm{v}_{\mu}.
\end{align}
Here, $b_{\mu}$ is equal to the $j$-th component of eigenvector $\bm{v}_{\mu}$. 
If $v_{\mu}(j)$ is the $j$-th component of $\bm{v}_{\mu}$, we obtain the equation of motion for the oscillation mode $a_{\mu}(\omega,\,t)$ as follows:
\begin{align}
  \frac{\partial^2 a_{\mu}(\omega,t)}{\partial t^2} \!+\! \gamma \frac{\partial a_{\mu}(\omega,t)}{\partial t}
  \!+\! \lambda_{\mu}\,a_{\mu}(\omega,t)\!=\! \sqrt{m_j}F\cos(\omega t)v_{\mu}(j).
  \label{F_O2}
\end{align}
The equation of motion (\ref{F_O2}) means that the oscillation dynamics on symmetrizable directed networks can be expressed by superposing the oscillations of each oscillation mode.

The solution of equation (\ref{F_O2}) is given by 
\begin{align}
  a_{\mu}(\omega,\,t) &= c_{\mu}{\rm e}^{-\frac{\gamma}{2}t}\cos\left(\sqrt{\lambda_{\mu} - \left(\frac{\gamma}{2}\right)^2}\,t + \phi_{\mu}\right) \nonumber\\
  &\quad + A_{\mu}(\omega)\cos(\omega t + \theta_{\mu}(\omega)),
  \label{ans}
\end{align}
where $c_{\mu}$ and $\phi_{\mu}$ are constants, and $A_{\mu}(\omega)$ and ${\theta}_{\mu}(\omega)$ are the amplitude and the initial phase, respectively.
They are expressed by using eigenfrequency $\omega_{\mu} = \sqrt{\lambda_{\mu}}$ as
\begin{align}
  A_{\mu}(\omega) &= \frac{\sqrt{m_j}\,F\,v_{\mu}(j)}{\sqrt{({{\omega}^2_{\mu}} - {\omega}^2)^2+({\gamma}\,{\omega})^2}},\label{amplitude}\\
  \theta_{\mu}(\omega) &= \arctan\left(-\frac{{\gamma}\,{\omega}}{{{\omega}^2_{\mu}} - {\omega}^2}\right).
  \label{initial_phase}
\end{align}
Since the first term on the right-hand side of the equation (\ref{ans}) attenuates with time, only the second term on the right-hand side remains after a long time.
Therefore, the stationary solution of equation (\ref{F_O2}) can be written as
\begin{align}
a_{\mu}(\omega,\,t) = A_{\mu}(\omega)\cos(\omega t + \theta_{\mu}(\omega)).
\label{ans2}
\end{align}
Accordingly, we obtain the stationary solution of the equation of motion (\ref{F_O_L0}) as follows:
\begin{align}
\bm{x}(t) = \bm{M}^{-\frac{1}{2}}\sum^{n-1}_{\mu = 0}A_{\mu}(\omega)\cos(\omega\,t + \theta_{\mu}(\omega))\,\bm{v}_{\mu}.
\end{align}

\subsection{Resonance on Networks}
\label{subsec.3C}
From (\ref{amplitude}), amplitude $A_{\mu}(\omega)$ takes maximal value at
\begin{align}
  \omega = \sqrt{\omega^2_{\mu}-\frac{\gamma^2}{2}}.
\end{align}
This phenomenon is called resonance.
If damping coefficient $\gamma$ is very small, 
amplitude $A_{\mu}(\omega)$ increases sharply around $\omega \simeq \omega_{\mu}$.

In practice, since amplitude $A_{\mu}(\omega)$ cannot be observed directly,
we focus on the oscillation energy in the stationary state. 
Oscillation energy $E_i(\omega)$ of node $i$ can be written as
\begin{align}
E_i(\omega) = \frac{1}{2}\,\omega^2\,\sum^{n-1}_{\mu = 0}(A_{\mu}(\omega))^2\,(v_{\mu}(i))^2.
\label{node_E}
\end{align}
The maximal values of $E_i(\omega)$ can be observed around $\omega \simeq \omega_{\mu}\,(\mu = 1,\,2,\ldots,\,n-1)$ except for minimum eigenfrequency ${\omega}_0 = 0$.

\section{Model of Online Flaming Phenomena Based on Resonance}
\label{sec.4}
In this section, based on the forced oscillation mentioned in Sec.~\ref{sec.3}, we propose a model of online flaming phenomena that is caused by the resonance created by a continuous external stimulus. 
Experiments on the oscillation energy of each node confirm the validity of the model.

\subsection{Concept of the Model}
\label{subsec.4A}
It has been revealed that the oscillation energy of each node represented by formula (\ref{node_E}) gives a generalized notion of node centrality, i.e., the significance of each node (user) \cite{takano,takano2}.
This means that oscillation phenomena on the graphs can describe the strength of the user's activity in the OSN of interest.
Therefore, we can regard the divergence of each node's oscillation energy as indicating the occurrence of online flaming.

Next, we consider the meaning of the damping coefficient, $\gamma$.
It is a parameter that describes the diminishment of user's interest in the information spread by mass media.
We assume that the content of the information determines $ \gamma$. 
The more newsworthy and attractive the information is, the smaller the value of $\gamma$ is.
Since it affects the strength and the abruptness of resonance, small $\gamma$ values are factors increasing the oscillation energy.

\begin{figure}[tb]
  \centering
  \includegraphics[width=9.0cm]{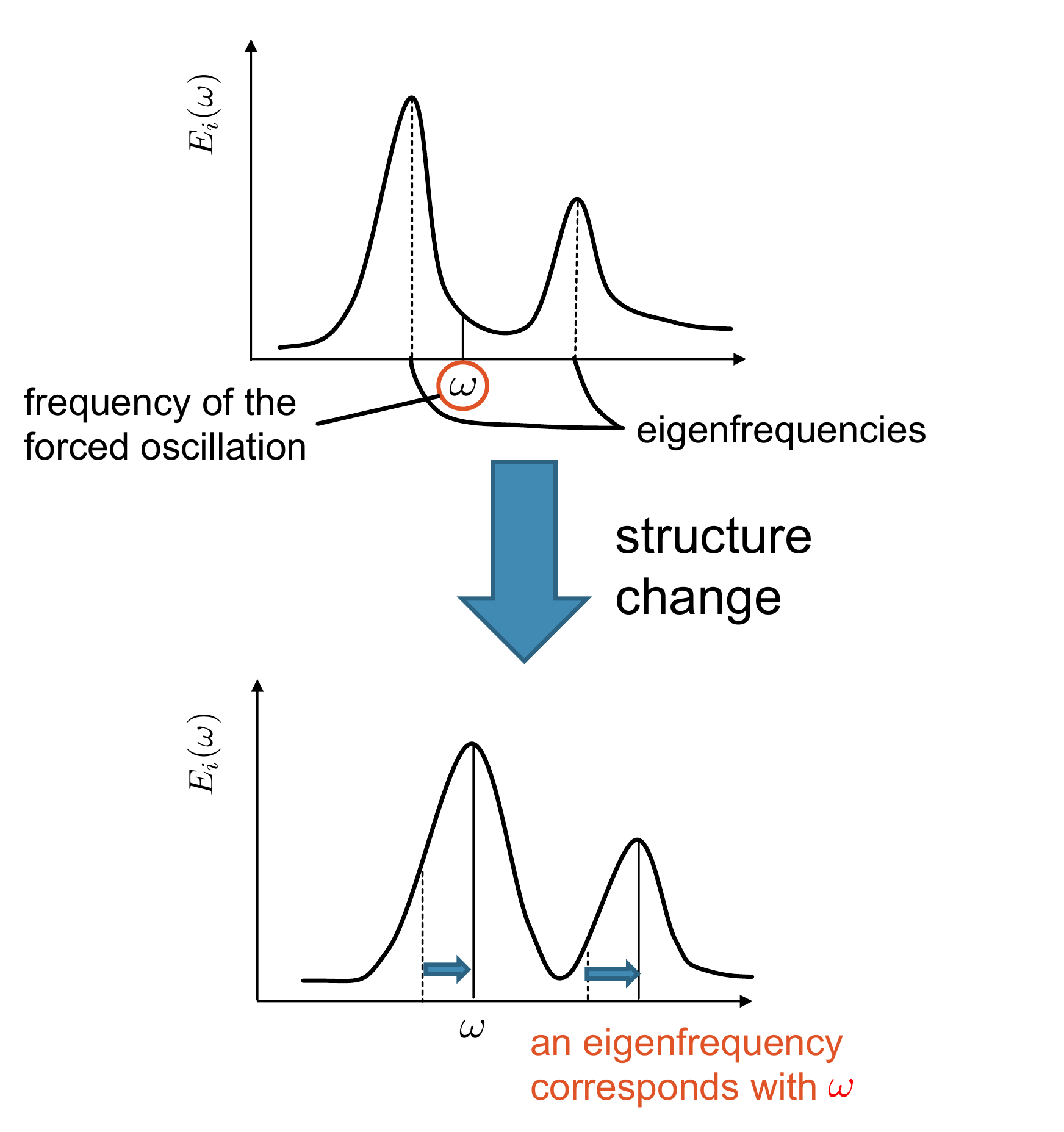}
  \caption{ Network structure change by resonance driven by external stimulus }
  \label{structure_change_eng}
\end{figure}

When mass media broadcast certain attractive information, user discussions about the information become heated.
To express this phenomenon in engineering terms, we consider that the network structure changes so that each node's oscillation energy increases.
That is, the network structure changes to resonate with a given external stimulus whose angular frequency is $\omega$, as shown in Fig.~\ref{structure_change_eng}.
For the original online social network, the eigenfrequency is $\omega_\mu = \sqrt{\lambda_\mu}$, where $\lambda_\mu$ $(\mu=0\,\,\dots,\,n-1)$ are the eigenvalues of the Laplacian matrix representing the network structure.
Let the maximum eigenfrequency satisfying $\omega_\nu \le \omega$ be $\omega_\nu$. 
We assume the structure of the network changes by $\omega_\nu \rightarrow \omega$. 
That is, $c := \omega/\omega_\nu$ is multiplied to the original  eigenfrequency $\omega_\nu$.
Also, all other eigenfrequencies are multiplied by $c$. 
Thus all the eigenvalues of the original Laplacian matrix are multiplied by $c^2$.
This corresponds to a change in which the weights of all links in the original OSN are multiplied by $c^2$.

Based on the above, we propose an online flaming model where each user's behavior is explosively activated by the resonance created by the release of topical information.

\subsection{Numerical Experiments on Online Flaming}
\label{subsec.4B}

\begin{figure}[tb]
    \centering
    \includegraphics[width=6.5cm]{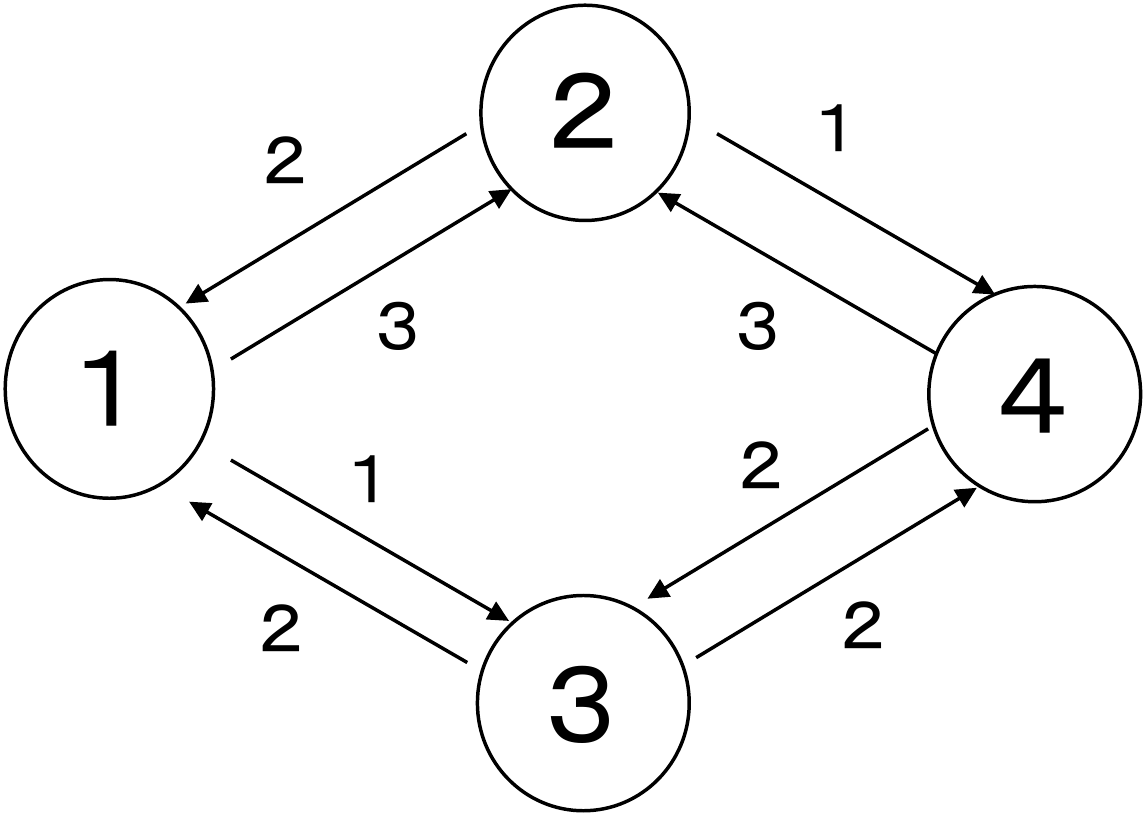}
    \caption{Symmetrizable directed graph with four nodes}
    \label{network_model_1}
\end{figure}

We use the symmetrizable directed graph with four nodes shown in Fig.~\ref{network_model_1} as the network model of interest.
Also, we arrange eigenfrequencies of its Laplacian matrix in ascending order as follows:
\begin{align}
(\omega_0,\,\omega_1,\,\omega_2,\,\omega_3) = (0,\,1.8288,\,2.1884,\,2.8047).
\end{align}
In this experiment, the amplitude $F$ of the external stimulus is 1, and we input the external stimulus to node 1.

\begin{figure}[tb]
  \centering
  \includegraphics[width=9.0cm,clip]{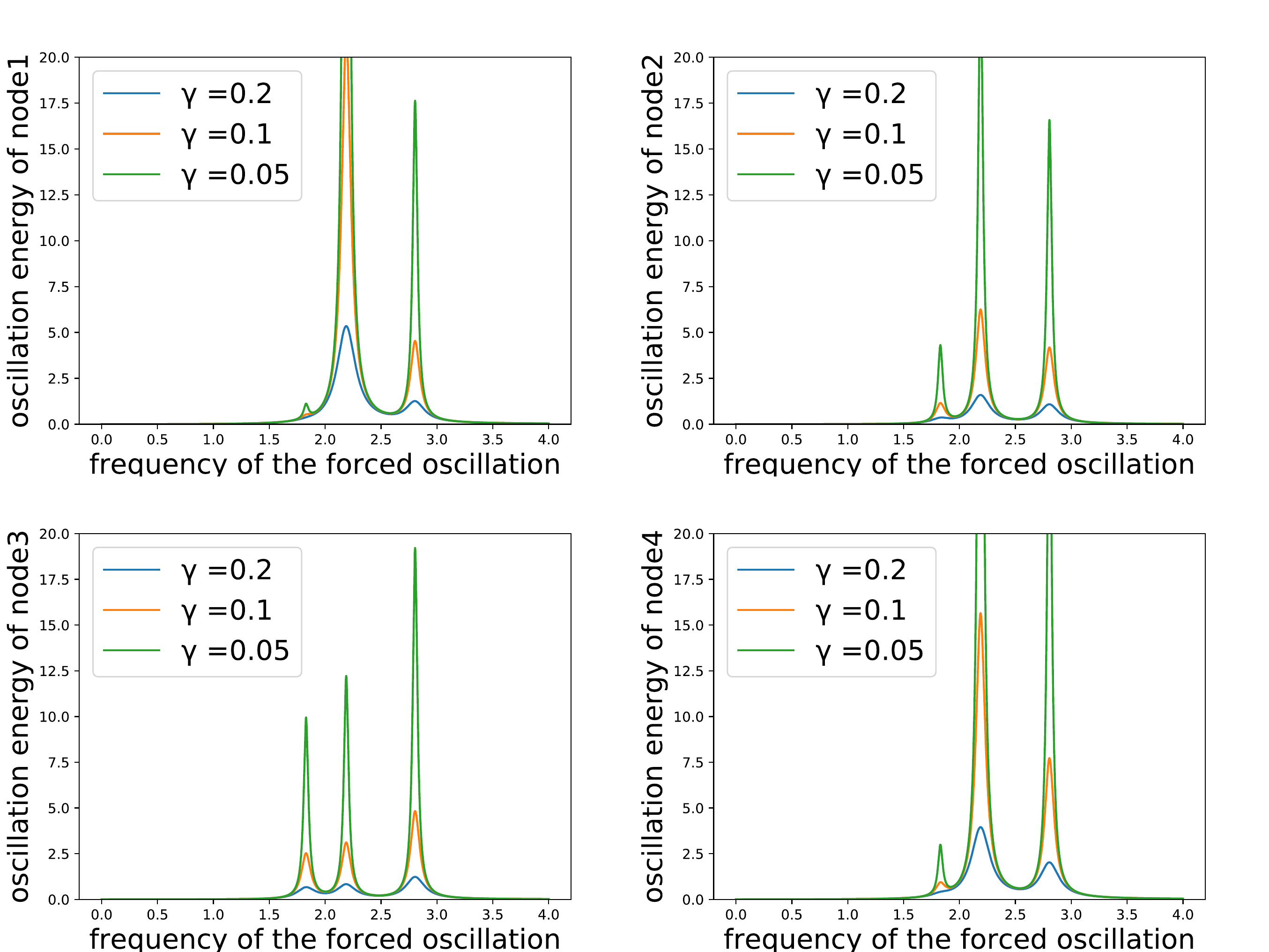}
  \caption{Oscillation energy of each node with respect to angular frequency $\omega$ of the forced oscillation }
  \label{oscillation_energy}
\end{figure}

Figure~\ref{oscillation_energy} shows the oscillation energy of each node for several damping coefficients, $\gamma$, and various angular frequencies, $\omega$, of the forced oscillation.
From these figures, we recognize that resonance occurs when $\omega$ is close to $\omega_{\mu}$ $(\mu = 1,\,2,\,3)$.
Also, the smaller damping coefficient $\gamma$ is, the higher the intensity of the resonance is.
Since the strength of resonance directly affects the magnitude of oscillation energy, this result shows that the network is explosively activated when $\omega \simeq \omega_{\mu}$ and $\gamma$ is small.
Thus, these numerical experiments show that explosive activation of user activity, such as online flaming, occurs even if all the Laplacian matrix's eigenvalues are real numbers.

By comparing these figures, it can be seen that the intensity of the resonance differs with the node and eigenfrequency.
This is because the absolute values of the components $v_{\mu}(i)$ and $v_{\mu}(j)$ of the eigenvector influence the oscillation energy of each node.
In particular, since $v_{\mu}(j)$ is included in the formula for the oscillation energy of any node, it significantly impacts the activation of the whole network.
Obviously, if the user receives information of interest, the strength of user dynamics for the whole OSN will be enhanced.

\section{Omen of Online Flaming}
\label{sec.5}
To minimize the damage caused by explosive dynamics such as online flaming, countermeasure(s) should be applied before flaming occurs.
Therefore, identifying the omen of online flaming is important.
This section shows that the appearance of low-frequency beats in the solution of the equation of motion can well predict online flaming; the validity of this approach is confirmed through a simulation model describing the oscillation on networks.

\subsection{Mechanism Underlying Beat Appearance}
\label{sec.5A}
We solve the equation of motion (\ref{F_O2}) for the oscillation mode under the following initial conditions:
\begin{align}
  a_{\mu}(\omega,\,0) = 0,\quad
  \frac{\partial a_{\mu}(\omega,\,t)}{\partial t}\biggl|_{t=0} = 0.
\end{align}
If the angular frequency $\omega$ is close to eigenfrequency $\omega_{\mu}$ and the damping coefficient $\gamma$ is very small, the constants $\phi_{\mu}$ and $c_{\mu}$ can be written as
\begin{align}
\phi_{\mu} = \pm \frac{\pi}{2}, \qquad c_{\mu} = \mp A_{\mu}(\omega),
\end{align}
respectively.
Therefore, the solution of (\ref{F_O2}) is given by
\begin{align}
a_{\mu}(\omega,\,t) = A_{\mu}(\omega)\,{\rm e}^{-\frac{\gamma}{2}t}\sin(\omega'_{\mu}\,t) -A_{\mu}(\omega)\sin(\omega\,t),
\end{align}
where $\omega'_{\mu} = \sqrt{\omega_{\mu}^2 - (\gamma/2)^2}$.
Moreover, in the case of small $t$, the solution can be rewritten by using the sum-to-product formula as
\begin{align}
a_{\mu}(\omega,\,t) \simeq 2A_{\mu}(\omega)\cos\left(\frac{\omega + \omega'_{\mu}}{2}\,t\right)\sin\left(\frac{\omega - \omega'_{\mu}}{2}\,t\right).
\label{beat}
\end{align}
Focusing on solution (\ref{beat}), since we consider the situation that the angular frequency $\omega$ is close to eigenfrequency $\omega_{\mu}$, the wave represented by the $\sin$ component has a low frequency.
Also, amplitude $A_{\mu}(\omega)$ is very large in case of $\omega \simeq \omega_{\mu}$.
Thus, the solution of the equation of motion yields the appearance of low-frequency beats and the increase in amplitude.

We associate both phenomena with the model of flaming mentioned in Sec.~\ref{sec.4}.
Eigenfrequency $\omega_{\mu}$ approaches angular frequency $\omega$ of forced oscillation as the network structure changes due to the external stimulus.
This means that a low-frequency beat will appear.
Therefore, the appearance of low-frequency beats can be used as an omen of online flaming.

\subsection{Numerical Experiments on Omen of Online Flaming}
\label{sec.5B}
To describe the behavior of each node, we introduce some node state variables. 
Let $v_i(\omega,\,t)$ be the time partial derivative of $x_i(\omega,\,t)$ at time $t$.
Based on the equation of motion (\ref{F_O_L0}), we can write the discrete-time temporal evolution equations, for the node receiving the external stimulus, of $v_i(\omega,\,t)$ and $x_i(\omega,\,t)$ as 
\begin{align}
v_i(\omega,t + \Delta t) &= v_i(\omega,t)
- \bigg(\gamma v_i(\omega,t) + \sum_{j \in \partial i}w_{ij}\Delta x_{ij}\bigg)\Delta t \notag \\
&\qquad {} + F{\rm cos}(\omega t)\Delta t,
\label{v_i}\\
x_i(\omega,t + \Delta t) &= x_i(\omega,t) + v_i(\omega,t)\Delta t,
\label{x_i}
\end{align}
respectively, where $\Delta x_{ij} = x_i(\omega,t) - x_j(\omega,t)$, and $\Delta t$ is a sufficiently small value.
Since the other nodes do not receive the external stimulus directly, their temporal evolution equation of $v_i(\omega,\,t)$ can be written as
\begin{align}
  v_i(\omega,t + \Delta t) &= v_i(\omega,t) - \bigg(\gamma v_i(\omega,t)
  + \sum_{j \in \partial i}w_{ij}\Delta x_{ij}\bigg)\Delta t,
\label{v_i_2}
\end{align}
and their $x_i(\omega,\,t)$ is the same as (\ref{x_i}).
Based on the above, we express their behaviors by applying (\ref{v_i}),~(\ref{x_i}) and (\ref{v_i_2}).

\begin{figure}[tb]
    \centering
    \includegraphics[width=5.5cm]{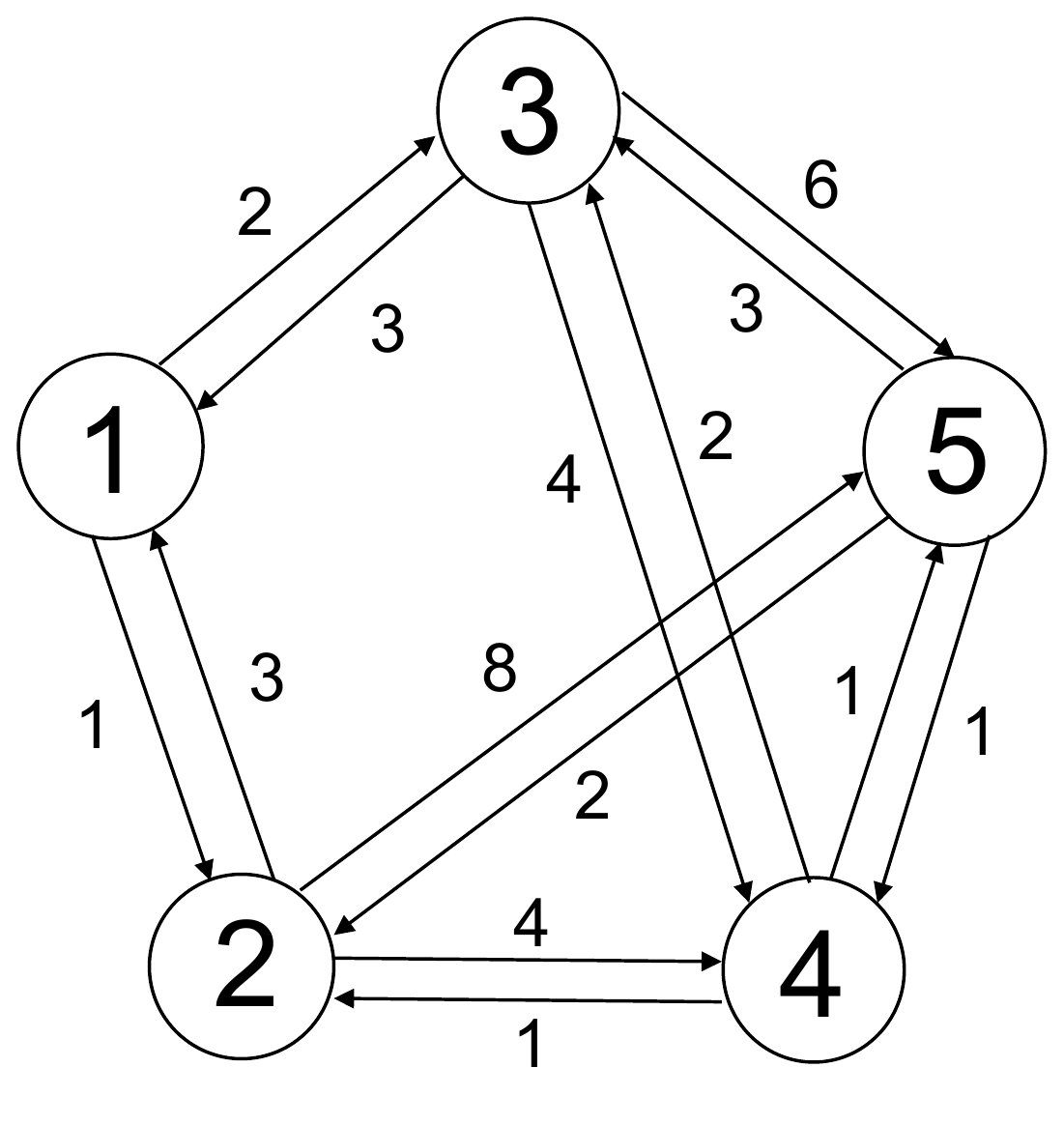}
    \caption{Symmetrizable directed graph with five nodes}
    \label{network_model_2}
\end{figure}

We use the symmetrizable directed graph with five nodes shown in Fig.~\ref{network_model_2} as the network model of interest. 
Also, we arrange the eigenfrequencies of its Laplacian matrix in ascending order as follows:
\begin{align}
(\omega_0,\omega_1,\omega_2,\omega_3,\omega_4) = (0,1.7734,2.3958,3.7417,4.2562).
\end{align}

In this simulation, we can input the external stimulus into node $1$ and derive the kinetic energy $K_i(\omega,\,t)$ of the other nodes, $i$, as
\begin{align}
K_i(\omega,\,t) = \frac{1}{2}\,m_i\,(v_i(\omega,\,t))^2.
\end{align}
Also, we take simple moving averages to more easily observe the beats.
Table~\ref{parameter} shows some parameter values used in this simulation.

\begin{table}[tb]
\centering
\caption{Parameter values}
\begin{tabular}{c|c}
parameters   & values     \\ \hline
$F$ & 1.0     \\ \hline
$\gamma$   & 0.02  \\ \hline
$\Delta t$ & 0.001 \\
\end{tabular}
\label{parameter}
\end{table}

\begin{figure}[tb]
  \centering
  \includegraphics[width=9.0cm,clip]{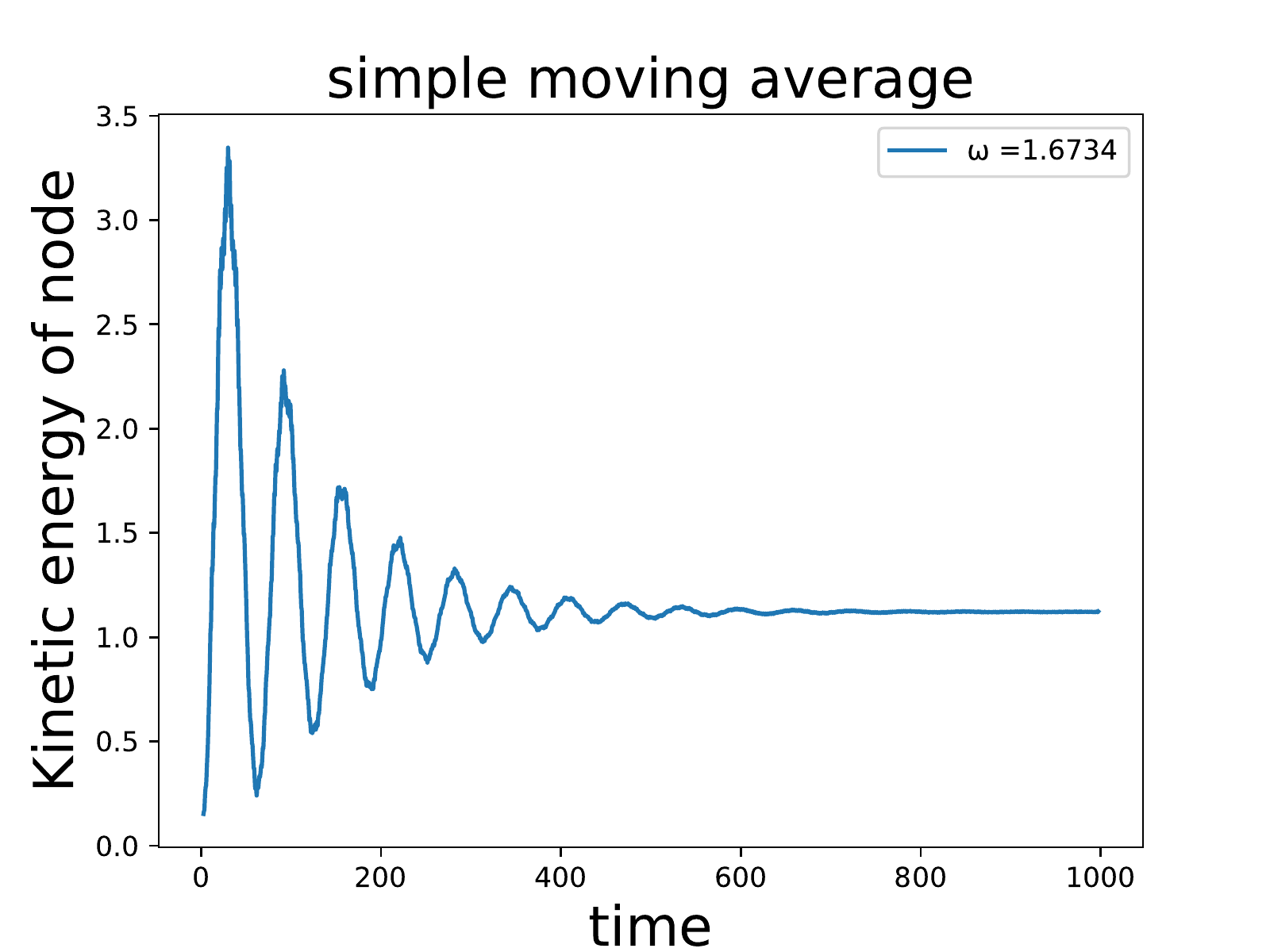}
  \caption{Simple moving average of the kinetic energy $K_i(\omega,\,t)$ with $\omega = \omega_1 - 0.1$}
  \label{kinetic_energy1}
\end{figure}

\begin{figure}[tb]
  \centering
  \includegraphics[width=9.0cm,clip]{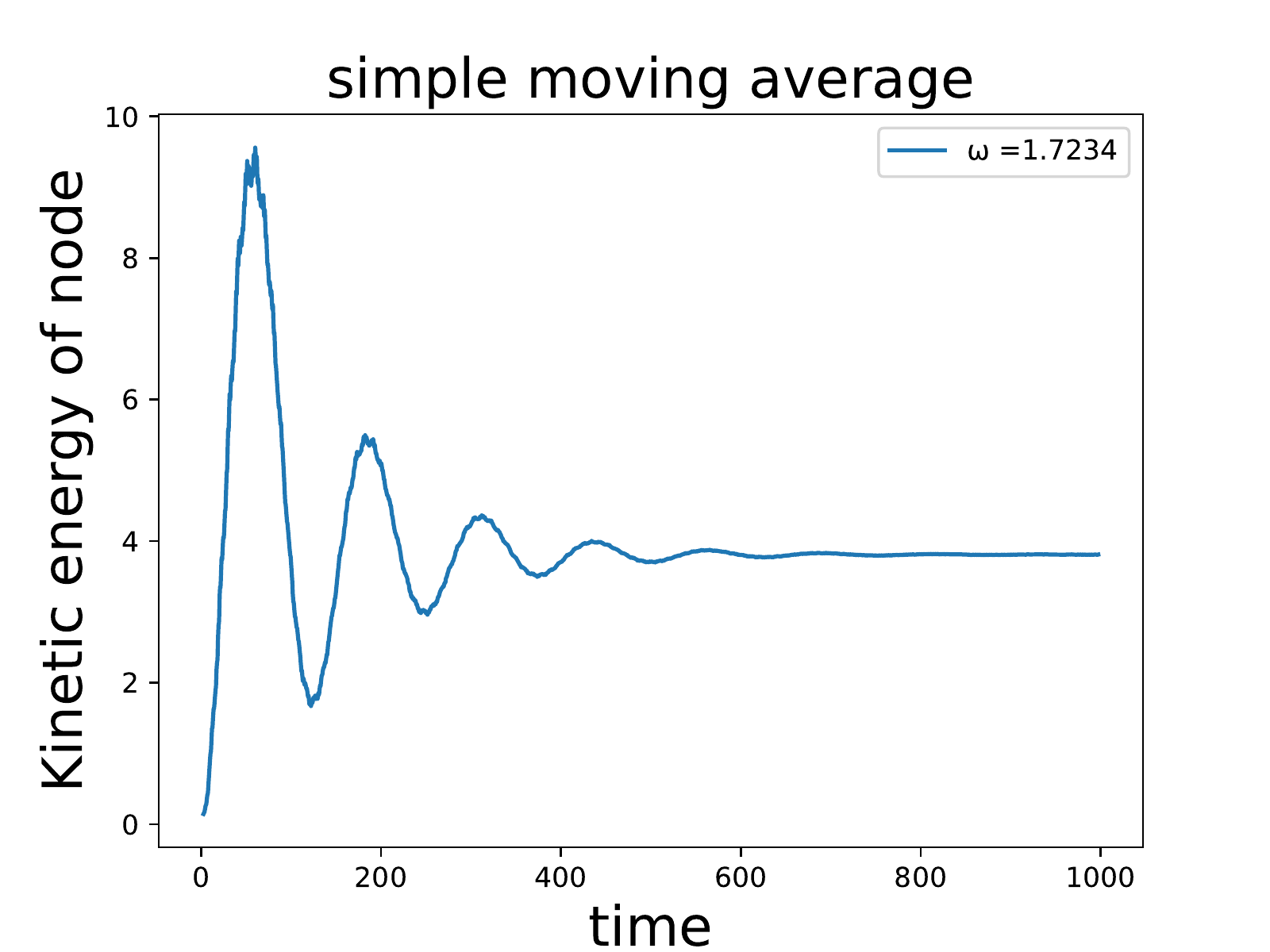}
  \caption{Simple moving average of the kinetic energy $K_i(\omega,\,t)$ with $\omega = \omega_1 - 0.05$}
  \label{kinetic_energy2}
\end{figure}

\begin{figure}[tb]
  \centering
  \includegraphics[width=9.0cm,clip]{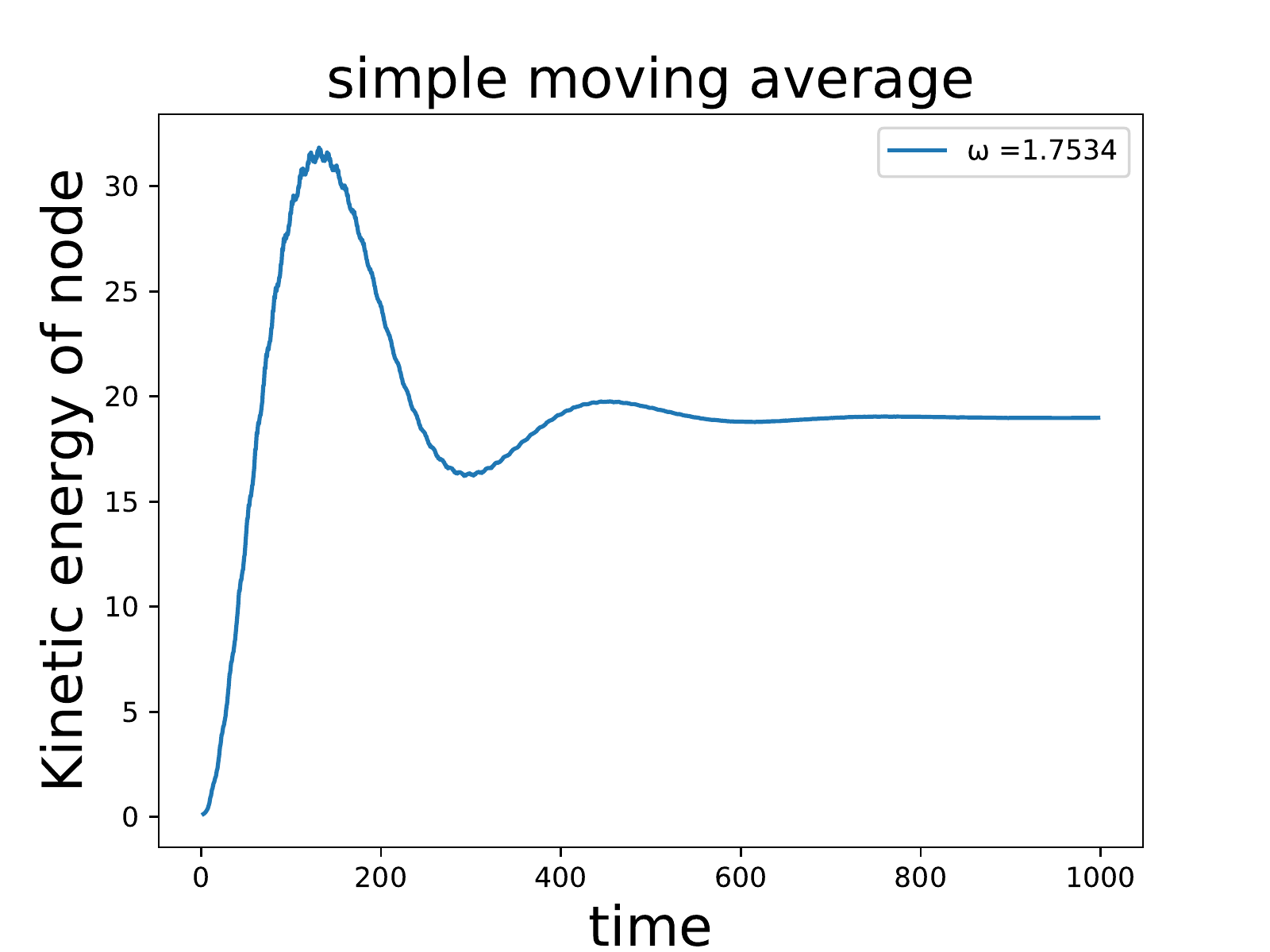}
  \caption{Simple moving average of the kinetic energy $K_i(\omega,\,t)$ with $\omega = \omega_1 - 0.02$}
  \label{kinetic_energy3}
\end{figure}

Figures~\ref{kinetic_energy1}, \ref{kinetic_energy2} and \ref{kinetic_energy3} plot the simple moving averages of the temporal change of kinetic energy $K_i(\omega,\,t)$ under the condition that $\omega = \omega_{1}-0.1,\,\omega_{1}-0.05,\,\omega_{1}-0.02$, respectively.
From these figures, it can be seen that the smaller the difference between $\omega$ and $\omega_{1}$ is, the lower the frequency of the beats is, and the larger the convergence value of simple moving averages of kinetic energy is.
Therefore, the results of the simulation confirm the validity of the theory described above.

Note that in this experiment, the beats appear only when $\omega \simeq \omega_{\mu}$.
As mentioned in Sec. \ref{sec.3}, the oscillation of each node on the network is the superposition of several oscillation modes, so the graph does not always show a well-shaped waveform.
It appears that the distorted graphs in Figs.~\ref{kinetic_energy1}, \ref{kinetic_energy2} and \ref{kinetic_energy3} represent waveform superposition.
However, since $\omega$ and $\omega_{\mu}$ are close to each other, the amplitude of the corresponding oscillation mode is substantial, and its influence becomes dominant.
Therefore, the beats can be observed.
On the other hand, as the difference between $\omega$ and $\omega_{\mu}$ increases, the effect of other oscillation modes becomes non-negligible, so it becomes more difficult to observe the beats.

The appearance of low-frequency beats, revealed by the theoretical framework and experiments in this section, is also observed by analyzing actual data.
In \cite{nagatani2}, the authors reported the increase of the low-frequency spectrum in the period having many posts by analyzing the logs of a Japanese major bulletin board system.
Thus, the theoretical framework described above can explain the omen of online flaming in the actual OSN. 
Note that a simple increase in the number of posts on a bulletin board system can not explain the low-frequency spectrum increase. 
The occurrence of low-frequency beat reinforces the validity of the oscillation model, which is underlying user dynamics in OSNs. 

\section{Conclusion}
\label{sec.6}
In this paper, we proposed a model of online flaming in social networks that describes how the resonance imposed by continuous external stimuli on the networks makes online flaming possible even if all the Laplacian matrix's eigenvalues are real numbers.
This is a new model that describes how explosive user dynamics can be triggered by the spread of information by mass media.
Also, we solved the equation of motion for the oscillation mode to provide a theoretical explanation of the appearance of low-frequency beats.
We consider that the emergence of such beats is a useful omen of online flaming that will enable flaming countermeasures to be implemented.
In the future, we plan to conduct simulations using complex and large-scale network models representing actual OSNs. 
Also, we will consider modeling the network structure changes created by resonance with external stimuli.

\section*{Acknowledgment}
This research was supported by Grant-in-Aid for Scientific Research (B) No.~19H04096 (2019--2021)  and No.~20H04179 (2020--2022) from the Japan Society for the Promotion of Science (JSPS).

\end{document}